\documentclass[12pt]{iopart}

\usepackage[utf8]{inputenc}
\expandafter\let\csname equation*\endcsname\relax
\expandafter\let\csname endequation*\endcsname\relax
\usepackage{amsmath}
\usepackage{float}
\usepackage{braket}
\usepackage[backend=biber,style=numeric,sorting=none]{biblatex}
\usepackage{graphicx}
\usepackage{hyperref}
\begin{document}

\title{The Quantum Eraser from a Weak Values Perspective}

\author{Tom Rivlin}

\address{Chemical and Biological Physics Department, Weizmann Institute of Science, 76100 Rehovot, Israel}
\vspace{10pt}
\begin{indented}
\item[]tom.rivlin@weizmann.ac.il
\item[]02 May 2022
\end{indented}

\begin{abstract}
The quantum eraser variant of the double-slit experiment, and its `delayed choice' sub-variant, are considered from the perspective of weak value and weak measurement theory (which is briefly reintroduced here). The interference fringes that appear when measuring certain spin states, which can then be `erased' when measuring other spin states, are shown to be anomalous weak values that depend on particular post-selection choices. By framing the choice of spin measurement as a weak value of a certain weak measurement, it is then made clear what physical claims can and cannot be made about what occurs in the quantum eraser experiment. Specifically, claims about the choice of spin-state `retrocausally' influencing the choice of slit(s) for the particles to travel through are discredited, and a simple framework is presented for understanding how the fringes arise and why they can be `erased'.
\end{abstract}

\vspace{2pc}
\submitto{\EJP}
%\maketitle

\section{Background}
The quantum eraser thought experiment is a simple, measurable variation on the double-slit experiment that is often called a quantum paradox, with non-intuitive implications \cite{82ScullyDruhl, 96Seager, 00KimYuKulik, 02WalbornCunhaPadua, 05AharonovZubairy, 15Ellerman, 19Carroll, 20Qureshi, 21BrackenHanceHossenfelder}. A naive interpretation of the `delayed choice' variant of the experiment can even be erroneously thought of an example of \textit{retrocausality} \cite{15Ellerman,21BrackenHanceHossenfelder}. The supposedly paradoxical nature of the experiment, however, is actually the result of a misinterpretation. The resolution to the supposed paradox has been discussed by several authors \cite{19Carroll, 20Qureshi, 21BrackenHanceHossenfelder}, but the resolution to the paradox has not previously been analysed from a weak measurement and weak value perspective (though see \cite{03Wiseman, 07MirLundeenMitchell, 11KocsisBravermanRavets, 13SteinbergFeizpourRozema, 13Svensson} for discussions of the related issue of weak measurement of \textit{momentum} in the two-slit setup). Here it will be shown that the experiment can be thought of as a type of weak measurement, similar to how other supposed quantum paradoxes can be examined from a weak measurement viewpoint \cite{16AharonovColomboPopescu,21KunstatterZiprick, 21CorreaSaldanha}. This will help to clarify what can and cannot be claimed about the experiment, and serves as an informative introduction to a genre of quantum theory problems that can be clarified with the weak values perspective.

The explanation for the setup (following the explanation of \cite{20Qureshi}) is as follows. Suppose one has a double-slit apparatus, such that a particle can pass through two narrow, finite-width slits $a$ and $b$ with small separation. In the typical quantum fashion, the particle's wavefunction can be written as a superposition of the states where it passes through $a$ and the states where it passes through $b$, and these states can interfere.
\begin{equation}
    \label{eq:psi1}
    \ket{\psi_1}=\alpha\ket{\psi_a}+\beta\ket{\psi_b}.
\end{equation}

If one were to observe the distribution of particles at a particular choice of $x$ at a screen on the other side of the slits (after a time $t$ has passed from the particle hitting the slits -- the $t$ variable should be considered implicit wherever an $x$ variable is imposed), one would see particles distributed according to:
\begin{equation}
    \label{eq:xpsi1sq}
    |\braket{x|\psi_1}|^2=|\alpha|^2|\psi_a(x)|^2+|\beta|^2|\psi_b(x)|^2+\alpha^*\beta\psi_a^*(x)\psi_b(x)+\beta^*\alpha\psi_b^*(x)\psi_a(x).
\end{equation}

The cross-terms in the final expression give rise to fringes caused by interference effects, as is well known from conventional quantum mechanics. For instance, following an example given in \cite{03Zettili}, if for some wavenumber $k$ and angular velocity $\omega$ at some time $t$ we have
\begin{equation}
    \label{eq:psiax}
    \begin{aligned}
    \psi_a(x)&=e^{-i\left(kx-\omega t\right)}/\sqrt{\pi\left(1+x^2\right)}\, , \\
    \psi_b(x)&=e^{-i\left(kx+\pi x-\omega t\right)}/\sqrt{\pi\left(1+x^2\right)}\, ,
    \end{aligned}
\end{equation}
this leads to the overall intensity distribution at the screen being given (for $\alpha=\beta=1/\sqrt{2}$) by
\begin{equation}
    \label{eq:xpsi1eg}
    |\braket{x|\psi_1}|^2=\frac{2\cos^2{\left(\pi x/2\right)}}{\pi\left(1+x^2\right)}.
\end{equation}
This expression gives a cosine squared distribution modulated by an envelope -- a standard interference fringe pattern. By contrast, if one were to plot $|\psi_a(x)|^2$ on its own, the distribution would have a single, distinct peak.

The `quantum eraser' variant is the situation where one couples each of the two states associated with the slits to one of two possible states of an observable, such as spin along the $z$-axis (or photon polarisation). This would be achieved by configuring the experiment in such a way that only spin-up particles (say) go through slit $a$ and spin-down through slit $b$. Then by measuring spin, we measure which slit was traversed (the ``which-way'' information). For orthogonal spin-up and down states $\ket{\uparrow}$ and $\ket{\downarrow}$, the wavefunction then looks like
\begin{equation}
    \label{eq:psi2}
    \ket{\psi_2}=\alpha\ket{\psi_a}\ket{\uparrow}+\beta\ket{\psi_b}\ket{\downarrow}.
\end{equation}
Note that this implies that there are two Hilbert spaces, and the overall wavefunction of the single system $\ket{\psi_2}$ is constructed by taking tensor products between states in the $\{\ket{\psi_a},\ket{\psi_b}\}$ Hilbert space and states in the $\{\ket{\uparrow},\ket{\downarrow}\}$ Hilbert space.

When one now observes the distribution of particles at a point $x$ on the screen, the distribution takes the form
\begin{equation}
    \label{eq:xpsi2sq}
    |\braket{x|\psi_2}|^2=|\alpha|^2|\psi_a(x)|^2+|\beta|^2|\psi_b(x)|^2,
\end{equation}
where the orthonormality of the two spin states eliminates the cross-terms in Eq.~\ref{eq:xpsi1sq}. Hence the slits are distinguished from the start, the interference effects are not present, and the final distribution will look like overlapping single-slit distributions. This is similar to the case where the interference is destroyed by a measurement that `looks at' the slits by shining photons on them and transferring momentum \cite{03Wiseman, 07MirLundeenMitchell}.

The `eraser' part of the quantum eraser comes when we consider the following: spin can be measured along any axis, so let us imagine measuring it along one orthogonal to the $z$-axis, in orthonormal states $\ket{\pm}=\left(\ket{\uparrow}\pm\ket{\downarrow}\right)/\sqrt{2}$. Because of the commutation relation between the two sets of spins, measuring it in this state `destroys' or `erases' the information about the specific up or down state it was in -- the which-way information. We can write $\ket{\psi_2}$ in terms of this new basis (and $\braket{x|\psi_2}$ will be the same as in Eq.~\ref{eq:xpsi2sq}):
\begin{equation}
    \label{eq:psi22}
    \ket{\psi_2}=\frac{1}{\sqrt{2}}\left(\alpha\ket{\psi_a}+\beta\ket{\psi_b}\right)\ket{+}+\frac{1}{\sqrt{2}}\left(\alpha\ket{\psi_a}-\beta\ket{\psi_b}\right)\ket{-}.
\end{equation}

The notable effects occur when one considers \textit{sub-ensembles} of $\ket{\psi_2}$. Suppose one has a detector that distinguishes particles at the screen as being in the $\ket{+}$ state or the $\ket{-}$ state. One then obtains either one of
\begin{equation}
\begin{aligned}
    \label{eq:pmpsi2sq}
    |\bra{+}\braket{x|\psi_2}|^2&=\frac{1}{2}\left(|\alpha|^2|\psi_a(x)|^2+|\beta|^2|\psi_b(x)|^2+\alpha^*\beta\psi_a^*(x)\psi_b(x)+\beta^*\alpha\psi_b^*(x)\psi_a(x)\right),\\
    |\bra{-}\braket{x|\psi_2}|^2&=\frac{1}{2}\left(|\alpha|^2|\psi_a(x)|^2+|\beta|^2|\psi_b(x)|^2-\alpha^*\beta\psi_a^*(x)\psi_b(x)-\beta^*\alpha\psi_b^*(x)\psi_a(x)\right).
\end{aligned}
\end{equation}

These expressions, of course, have interference terms. Contrast that with what happens when we distinguish particles at the screen as being in the $\ket{\uparrow}$ or $\ket{\downarrow}$ states:
\begin{equation}
\begin{aligned}
    \label{eq:udpsi2sq}
    |\bra{\uparrow}\braket{x|\psi_2}|^2&=|\alpha|^2|\psi_a(x)|^2,\\
    |\bra{\downarrow}\braket{x|\psi_2}|^2&=|\beta|^2|\psi_b(x)|^2.
\end{aligned}
\end{equation}
When we distinguish parts of the distribution in this way, the particle is only ever detected as having been through slit $a$ for the $\ket{\uparrow}$ case or slit $b$ for the $\ket{\downarrow}$ case.

So, depending on which axis we \textit{choose} to measure the spin along, we either do or do not see interference fringes. This has been realised experimentally (with photon polarisations) \cite{00KimYuKulik, 02WalbornCunhaPadua}, and, as expected, interference fringes were seen for one choice of sub-ensembles, but not for the other.

Of course, one is free to choose which axis to measure the spin along \textit{after} the particles have traveled through the slit(s) -- this is where the `delayed choice' name comes from. One could think of this as being ``retrocausal'' -- that the choice one makes in the present as to which axis to measure spin along causes the particle in the past to either go through two slits or one -- to interfere or not interfere. Different sub-ensembles being revealed by interactions between the system and the environment is a well-known phenomenon \cite{00EnglertBergou}. Here the claim, however, is that the experimenter's choice after the measurement can alter the sub-ensembles.

The claim is untrue for a number of reasons \cite{15Ellerman,21BrackenHanceHossenfelder}, and the rest of this discussion will focus on how to understand this experiment using weak measurement theory. But first it must be noted that the interference patterns one observes by measuring along the $\pm$ axis are \textit{not} the same as the patterns one observes by measuring state $\ket{\psi_1}$, the state without any attached spins. The $|\bra{+}\braket{x|\psi_2}|^2$ state's values are half those of $|\braket{x|\psi_1}|^2$, and the structure of $|\bra{-}\braket{x|\psi_2}|^2$ is a modified version of the $|\braket{x|\psi_1}|^2$ pattern due to the minus signs in the cross-terms. Measuring the spin along the $\pm$ axis still changes the measurement outcome from what one would observe without measuring any spin, just not as drastically as measuring along the up-down axis. And, of course, if one were to add the two $\pm$ or up-down sub-ensembles together, one would exactly retrieve the original $|\braket{x|\psi_2}|^2$ distribution with no interference. 

\section{Weak Measurement and Weak Values}

Weak measurement theory \cite{91StoryRitchieHulet, 08AharonovVaidman, 13TamirCohen, 14AharonovCohenElitzur} offers an interesting perspective on this problem, and will be briefly introduced here (based on \cite{13TamirCohen}) in a form that facilitates its application to the quantum eraser. Consider an attempt to obtain an eigenvalue of an observable $\hat{A}$ of a quantum system $\ket{\phi_i}$, \textit{conditional} on the system being \textit{post-selected} to be in state $\ket{\phi_f}$ \cite{95Steinberg}. The \textit{weak value} of the observable for this post-selection is then said to be
\begin{equation}
    \label{eq:WV}
    \braket{\hat{A}}_{w,i,f}=\frac{\bra{\phi_f}\hat{A}\ket{\phi_i}}{\braket{\phi_f|\phi_i}}.
\end{equation}

One can then sum over all the possible weak values -- all the possible post-selected states $\ket{\phi_f}$ -- multiplied by the probabilities of observing each value. This will give the standard $\braket{\hat{A}}$ for the quantum system $\ket{\phi_i}$. This is simple Bayesian logic \cite{95Steinberg, 10HosoyaShikano}, but the caveat is that the weak values of $\hat{A}$ may in general be complex, unlike the strong values which are guaranteed to be real. Even when not complex, these values may be highly \textit{anomalous} -- falling well outside of the regular eigenvalue spectrum of the observable.

This is connected to the idea of weak measurement: a type of measurement which only perturbs the system it measures by a small amount. Consider a detector for the observable $\hat{A}$ modeled as a `pointer' coupled to the observable by an interaction Hamiltonian \cite{13TamirCohen}:
\begin{equation}
    \label{eq:Hint}
    \hat{H}_{\rm{int}}=g \hat{A}\otimes\hat{P}_d,
\end{equation}
where $\hat{P}_d$ is the operator representing the momentum of the pointer, and $g$ is a coupling parameter which is assumed to be small. This implies that interacting with the system being measured via this Hamiltonian gives the pointer a `kick', and alters its position, but only by a small amount, and with a large variance in the final position. With certain assumptions (e.g. different pointer positions correspond to different eigenvalues), this can be shown to translate into a large uncertainty about the eigenvalue of $\hat{A}$ that is being `pointed' to by the detector. Unlike with regular `strong' (or `von Neumann') measurements, the system being measured is still in a superposition of multiple eigenvalues of $\hat{A}$ after the interaction with the pointer.

Expanding to first order the time-evolution operator up to time $t$, $\exp{\left(-i \hat{H}_{\rm{int}} t/\hbar\right)}$, sandwiched between the pre- and post-selected states $\ket{\phi_i}$ and $\ket{\phi_f}$, allows one to derive the same expression as in Eq.~\ref{eq:WV} for the `expected' value of $\hat{A}$. This shows that weak values are the outcomes of weak measurements. In other words, after the weak measurement, the system is still in a superposition of multiple eigenstates of the observable, and the `probability' of observing any one of those eigenvalues -- of the system being in that post-selected state -- is given by the weak value associated with it. Strictly speaking, this is not actually a probability, but something known as a `quasiprobability', since it may be complex. (Weak values can also be obtained from sequences of strong measurements \cite{18CohenPollak}, meaning weak values and weak measurements do not always have to be considered in tandem.) It is notable that the form of the weak value can be derived both from weak measurements and from considering conditional probabilities.

\section{Weak Values of the Quantum Eraser}

Returning to the quantum eraser experiment, consider performing for the $\ket{\psi_2}$ system a weak measurement of the position with the operator $\hat{O}_x=\ket{x}\bra{x}$, which is a projection onto the $\ket{x}\bra{x}$ subspace and corresponds to a measurement of the density at position $x$. Here we also implicitly assume that included in this operator is the fact that time is propagated forward by some amount $t$, such that the waves can travel from the slits to the screen. Furthermore, strictly speaking the full operator is a tensor product of $\hat{O}_x$ and an identity operator in the Hilbert space of the spins -- the $\hat{O}_x$ operator only modifies states in the $\{\ket{\psi_a},\ket{\psi_b}\}$ Hilbert space.

Consider first the weak value of the operator for the post-selected state $\ket{\uparrow}$ (again, with an implicit identity operator in the other Hilbert space). It is given by
\begin{equation}
    \label{eq:upWV}
    \braket{\hat{O}}_{x,w,\uparrow}=\frac{\braket{\uparrow|x}\braket{x|\psi_2}}{\braket{\uparrow|\psi_2}}=\frac{\braket{\psi_2|\uparrow}\braket{\uparrow|x}\braket{x|\psi_2}}{\braket{\psi_2|\uparrow}\braket{\uparrow|\psi_2}}=\frac{|\alpha|^2\braket{\psi_a|x}\braket{x|\psi_a}}{|\alpha|^2\braket{\psi_a|\psi_a}}=|\psi_a(x)|^2.
\end{equation}
Likewise, the weak value of the position post-selected for the $\ket{\downarrow}$ state is given by $|\psi_b(x)|^2$. In other words, post-selecting for outcomes of the quantum evolution of $\ket{\psi_2}$ such that spin-up (down) is observed will only show you the parts of the distribution consistent with going through slit $a$ ($b$), so interference is eliminated. This is to be expected from the form of Eq.~\ref{eq:udpsi2sq}. Note that these weak values are independent of $\alpha$ and $\beta$.

The probability of observing each of the two post-selected states is
\begin{equation}
    \label{eq:PaPbupdown}
    \begin{aligned}
        \braket{\psi_2|\uparrow}\braket{\uparrow|\psi_2}=|\alpha|^2,\\
        \braket{\psi_2|\downarrow}\braket{\downarrow|\psi_2}=|\beta|^2,\\        
    \end{aligned}
\end{equation}
and so the expected value of the $x$-projection operator acting on $\ket{\psi_2}$ is obtained by summing over the two weak values multiplied by their probabilities, as expected. The result is the same as Eq.~\ref{eq:xpsi2sq}.

A similar argument works for post-selecting one of the $\ket{\pm}$ states when observing position to calculate $\braket{\hat{O}}_{x,w,+}$ and $\braket{\hat{O}}_{x,w,-}$:
\begin{equation}
    \label{eq:pWV}
    \frac{\braket{\pm|x}\braket{x|\psi_2}}{\braket{\pm|\psi_2}}=\left(|\alpha|^2|\psi_a(x)|^2+|\beta|^2|\psi_b(x)|^2\pm\alpha^*\beta\psi_a^*(x)\psi_b(x)\pm\beta^*\alpha\psi_b^*(x)\psi_a(x)\right),
\end{equation}
since $\braket{\psi_a|\psi_b}=0$, $\braket{\psi_a|\psi_a}=\braket{\psi_b|\psi_b}=1$, $|\alpha|^2+|\beta|^2=1$.

Hence we can clearly conclude that one would observe interference fringes in the $x$-distribution when post-selecting for one of the $\ket{\pm}$ states. With the up and down states, by contrast, no fringes would be present.

The probabilities for the $\ket{\pm}$ states reduce down to
\begin{equation}
    \label{eq:PaPbpm}
    \braket{\psi_2|\pm}\braket{\pm|\psi_2}=\frac{1}{2},
\end{equation}
such that the same value as in Eq.~\ref{eq:xpsi2sq} is obtained by the appropriate sum over weak values multiplied by probabilities. Note that these probabilities are independent of $\alpha$ and $\beta$, and in the simplest case where $\alpha=\beta=1/\sqrt{2}$, the probabilities of observing the $\ket{\uparrow}$, $\ket{\downarrow}$, $\ket{+}$, and $\ket{-}$ states are the same.

\begin{figure}[ht]
    \centering
    \includegraphics[scale=0.5]{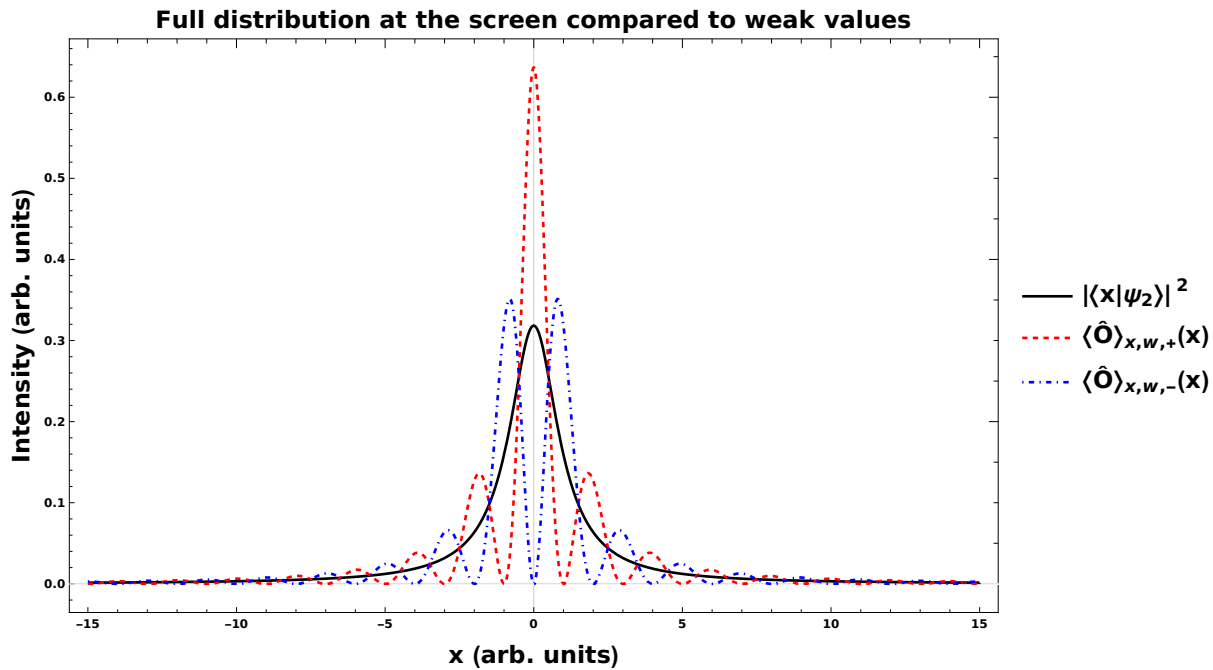}
    \caption{The actual distribution at the screen for $\psi_2$ compared with the $\ket{\pm}$ weak values (the $\uparrow$ and $\downarrow$ weak values exactly overlap the actual distribution, and the $|\braket{x|\psi_1}|^2$ distribution exactly overlaps the $\braket{\hat{O}}_{x,w,+}$ distribution). At every value of $x$, one of the two weak values is larger than the actual observed intensity.}
    \label{fig:Figure1}
\end{figure}

Of course, it is entirely possible for the weak values $\braket{\hat{O}}_{x,w,\uparrow}$, $\braket{\hat{O}}_{x,w,\downarrow}$, $\braket{\hat{O}}_{x,w,+}$, or $\braket{\hat{O}}_{x,w,-}$ to be larger than the value of $|\braket{x|\psi_2}|^2$ at certain values of $x$. This can be seen in Figure \ref{fig:Figure1}, which plots the actual intensity and weak values for the $\ket{\pm}$ states for the distributions in Equation \ref{eq:psiax} (with $\alpha=\beta=1/\sqrt{2}$). At each value of $x$, one of the two weak values $\braket{\hat{O}}_{x,w,+}$, $\braket{\hat{O}}_{x,w,-}$ is larger than $|\braket{x|\psi_2}|^2$. 

By contrast, the weak values $\braket{\hat{O}}_{x,w,\uparrow}$ and $\braket{\hat{O}}_{x,w,\downarrow}$, if plotted for each $x$, would both precisely overlap $|\braket{x|\psi_2}|^2$. Since the patterns of the two slits no longer interfere with each other, and since each of $\ket{\uparrow}$ and $\ket{\downarrow}$ are coupled to precisely one of the two slits, post-selecting for these states does not change the overall distribution.

At any given $x$, there is a greater chance of observing one of the $\ket{\pm}$ sub-ensembles of the overall distribution than there is of observing the distribution itself. Such phenomena are commonly seen with weak values. For instance, at $x=0$, the intensity of the $\braket{\hat{O}}_{x,w,+}$ distribution is twice that of the regular distribution. The explanation for this is clear: the probability of the particle being in the $\braket{\hat{O}}_{x,w,+}$ distribution is exactly half, and at that point the intensity of the $\braket{\hat{O}}_{x,w,-}$ distribution is 0. It is clear, however, that for all four weak value distributions considered here, there will be no negative weak values, and the probabilities of observing each of the weak values are  strictly positive and less than $1$.

\section{Discussion}

The weak measurement being performed is that of the density of the wavefunction at $x$, via the operator $\hat{O}_x$. The assumption is that because the measurement is weak, only a small amount of information about the distribution at the screen is obtained from it, and thus this observable still has a distribution of possible values. One is then free to choose to partition this distribution into sub-ensembles using different weak values. 
%Also the up or down (or plus or minus) spin state is correlated with position in $\ket{\psi_2}$, but because the measurement is weak, the system remains in a superposition of up or down (or plus or minus) after the measurement. The probability of it being measured in the up or down (or plus or minus) state is then given by the appropriate weak value. For the $\ket{\pm}$ spin states, the weak values have fringe patterns as functions of $x$, and for the $\ket{\uparrow}$, $\ket{\downarrow}$ states, the weak values do not.

Whether one subdivides the ensemble using a pair of up and down spins, or a pair of plus and minus spins, the overall probability of observing an electron at the point $x$ does not change. Once the slits are coupled to certain spin states, as they are here, the interference fringes will disappear, as expected. The post-selection choice does not `retrocausally' induce or destroy a fringe pattern in the spatial distribution of the wavefunction. In fact, nothing physical about the system is altered by the post-selection. 

Within a given post-selection choice, it is possible to obtain different probabilities, and here it is shown that it is possible to reveal a fringe pattern from the overall data set by post-selecting an appropriate weak value to measure. These weak values may even be `anomalous' in that the probability of observing the particle at a position $x$ \textit{conditional} on it being in (say) the $\ket{+}$ state may be larger than the total probability of observing it there. The correct sum over weak values, however, will always produce the true, physical, fringeless distribution one expects when summing over all possibilities, regardless of whether those weak values have fringes.

Also note that the operator only changes one of the two Hilbert spaces of the system, and the post-selection bra only targets the other of the two Hilbert spaces. Both of these Hilbert spaces are assumed to interact with the Hilbert space of the measuring device as per the derivation of the weak value.

Notably, in the experimental realization of the quantum eraser \cite{00KimYuKulik}, the two single-slit distributions at the screen significantly overlapped. In these circumstances, the which-way information gained from performing a measurement of position for a given particle is small, and that is true whether there is a fringed distribution or not. One can pick out a pattern with fringes by subdividing the overlapping distributions of particle detection events in an appropriate way as described by the weak values.

The mystery of the disappearing fringes in the quantum eraser thought experiment is nothing more than the result of subdividing a distribution into particular choices of sub-ensembles. As such, the analysis is amenable to an interpretation based on weak values and weak measurements, as shown here. The different spin measurements are shown to be different post-selected states of a weak measurement of the position, and the appropriate choice of post-selected state is shown to induce fringes in the distribution despite them not being present in the overall distribution. The physical implications of these findings are discussed, and it is explained how the supposed `retrocausality' of the delayed-choice quantum eraser is illusory.

\noindent\textbf{Acknowledgements}: This work was supported by Grant No. 408/19 of the Israel Science Foundation. I also wish to thank Prof. Eli Pollak and Dr. Eli Cohen for many useful discussions in regards to this work. 

\printbibliography

\end{document}